\documentclass[12pt,psfig]{article}
\newtheorem{stat}{Statement}[section]
\newcommand{\bstat}{\begin{stat}}
\newcommand{\estat}{\end{stat}}
\makeatletter
\def\section{\@startsection {section}{1}{\z@}{-3.5ex plus -1ex minus
 -.2ex}{2.3ex plus .2ex}{\large\bf}}
\def\subsection{\@startsection{subsection}{2}{\z@}{-3.25ex plus -1ex 
minus -.2ex}{1.5ex plus .2ex}{\normalsize\bf}}
\makeatother
\makeatletter

\@addtoreset{equation}{section}

\makeatother

\newsavebox{\uuunit}
\sbox{\uuunit}
    {\setlength{\unitlength}{0.825em}
     \begin{picture}(0.6,0.7)
        \thinlines
        \put(0,0){\line(1,0){0.5}}
        \put(0.15,0){\line(0,1){0.7}}
        \put(0.35,0){\line(0,1){0.8}}
       \multiput(0.3,0.8)(-0.04,-0.02){12}{\rule{0.5pt}{0.5pt}}
     \end {picture}}

\def\IP{\relax{\rm I\kern-.18em P}}

\begin{document}

\font\cmss=cmss10 \font\cmsss=cmss10 at 7pt
\def\twomat#1#2#3#4{\left(\matrix{#1 & #2 \cr #3 & #4}\right)}
\def\inbar{\vrule height1.5ex width.4pt depth0pt}
\def\IC{\relax\,\hbox{$\inbar\kern-.3em{\rm C}$}}
\def\IG{\relax\,\hbox{$\inbar\kern-.3em{\rm G}$}}
\def\IB{\relax{\rm I\kern-.18em B}}
\def\ID{\relax{\rm I\kern-.18em D}}
\def\IL{\relax{\rm I\kern-.18em L}}
\def\IF{\relax{\rm I\kern-.18em F}}
\def\IH{\relax{\rm I\kern-.18em H}}
\def\II{\relax{\rm I\kern-.17em I}}
\def\IN{\relax{\rm I\kern-.18em N}}
\def\IP{\relax{\rm I\kern-.18em P}}
\def\IQ{\relax\,\hbox{$\inbar\kern-.3em{\rm Q}$}}
\def\bfzero{\relax\,\hbox{$\inbar\kern-.3em{\rm 0}$}}
\def\IK{\relax{\rm I\kern-.18em K}}
\def\IG{\relax\,\hbox{$\inbar\kern-.3em{\rm G}$}}
 \font\cmss=cmss10 \font\cmsss=cmss10 at 7pt
\def\IR{\relax{\rm I\kern-.18em R}}
\def\ZZ{\relax\ifmmode\mathchoice
{\hbox{\cmss Z\kern-.4em Z}}{\hbox{\cmss Z\kern-.4em Z}}
{\lower.9pt\hbox{\cmsss Z\kern-.4em Z}}
{\lower1.2pt\hbox{\cmsss Z\kern-.4em Z}}\else{\cmss Z\kern-.4em
Z}\fi}
\def\bfone{\relax{\rm 1\kern-.35em 1}}
\def\dop{{\rm d}\hskip -1pt}
\def\real{{\rm Re}\hskip 1pt}
\def\trace{{\rm Tr}\hskip 1pt}
\def\ii{{\rm i}}
\def\diag{{\rm diag}}
\def\sch#1#2{\{#1;#2\}}
\def\bfone{\relax{\rm 1\kern-.35em 1}}
\font\cmss=cmss10 \font\cmsss=cmss10 at 7pt
\def\a{\alpha} \def\b{\beta} \def\d{\delta}
\def\e{\epsilon} \def\c{\gamma}
\def\G{\Gamma} \def\l{\lambda}
\def\L{\Lambda} \def\s{\sigma}
\def\cA{{\cal A}} \def\cB{{\cal B}}
\def\cC{{\cal C}} \def\cD{{\cal D}}
\def\cF{{\cal F}} \def\cG{{\cal G}}
\def\cH{{\cal H}} \def\cI{{\cal I}}
\def\cJ{{\cal J}} \def\cK{{\cal K}}
\def\cL{{\cal L}} \def\cM{{\cal M}}
\def\cN{{\cal N}} \def\cO{{\cal O}}
\def\cP{{\cal P}} \def\cQ{{\cal Q}}
\def\cR{{\cal R}} \def\cV{{\cal V}}\def\cW{{\cal W}}
\newcommand{\be}{\begin{equation}}
\newcommand{\ee}{\end{equation}}
\newcommand{\bea}{\begin{eqnarray}}
\newcommand{\eea}{\end{eqnarray}}
\let\la=\label \let\ci=\cite \let\re=\ref
%
%
%
\def\crr{\crcr\noalign{\vskip {8.3333pt}}}
\def\tilde{\widetilde}
\def\bar{\overline}
\def\us#1{\underline{#1}}
\def\IE{\relax{{\rm I\kern-.18em E}}}
\def\cE{{\cal E}}
\def\rt{{\cR^{(3)}}}
\def\IGam{\relax{{\rm I}\kern-.18em \Gamma}}
\def\IGa{\IA}
\def\ii{{\rm i}}
\def\beq{\begin{equation}}
\def\eeq{\end{equation}}
\def\beqa{\begin{eqnarray}}
\def\eeqa{\end{eqnarray}}
\def\nn{\nonumber}

\begin{titlepage}
\setcounter{page}{0}

\begin{flushright}

SISSA REF 55/99/EP

SWAT/227

\end{flushright}

\vskip 26pt

\begin{center}

{\Large \bf The generating solution of regular N=8 BPS black holes}

\vskip 20pt

{\large M. Bertolini$^a$, P. Fr\`e$^b$ and M. Trigiante$^c$}

\vskip 20pt

{\it $^a$International School for Advanced Studies ISAS-SISSA and INFN \\
Sezione di Trieste, Via Beirut 2-4, 34013 Trieste, Italy}

\vskip 5pt

{\it $^b$Dipartimento di Fisica Teorica, Universit\`a di Torino and INFN \\
Sezione di Torino, Via P. Giuria 1, 10125 Torino, Italy}

{\it $^c$Department of Physics, University of Wales Swansea, Singleton Park \\
Swansea SA2 8PP, United Kingdom}
\end{center}

\begin{abstract}
In this paper we construct the 5 parameter {\it generating solution} of $N=8$ BPS regular supergravity 
black holes as a five parameter solution of the $N=2$ $STU$ model. Our solution has a simpler form 
with respect to previous constructions already appeared in the literature and moreover, through the 
embedding $[SL(2)]^3\subset SU(3,3)\subset E_{7(7)}$ discussed in previous papers, the action of the 
$U$--duality group is well defined. This allows to reproduce via $U$--duality rotations any other 
solution, like those corresponding to R--R black holes whose microscopic description is given by 
intersecting D--branes.

\end{abstract}

\vskip 150pt

\begin{flushleft}
{\footnotesize
e-mail: teobert@sissa.it, fre@to.infn.it, m.trigiante@swansea.ac.uk}
\end{flushleft}

\vspace{2mm} \vfill \hrule width 3.cm
\vskip 0.2cm
{\footnotesize
Supported in part by   EEC  under TMR contracts ERBFMRX--CT96--0045 and
ERBFMRX--CT96--0012}

\vskip 20pt
\end{titlepage}
\section{Introduction}
\label{introgen}

After the advent of D-branes \cite{pol} there has been a renewed interest in the study of
supergravity black $p$-branes, in particular those preserving a fraction of the original 
supersymmetry. This is due to their identification with the BPS saturated non--perturbative states
of superstring theory \cite{duffrep,kstellec} which has promoted them from classical solutions of the
low--energy theory to solutions of the whole quantum theory. Therefore they represent an important
tool in probing the non--perturbative regime of superstring theories. Of particular interest are the
regular ones, namely those having a non--vanishing Bekenstein--Hawking entropy.

In this paper we deal with BPS static black hole solutions of $D=4$
supergravity preserving 1/8 of the original $N=8$ supersymmetry,
completing a program started in \cite{mp1} and  continued in
\cite{bft}. Let us recall that in the context of toroidally
compactified type II  supergravity the only regular black hole
solutions are the 1/8 supersymmetry preserving ones. The  1/2 and 1/4
black holes, whose general form has been completely classified in
\cite{mp2}, have a  vanishing horizon area.  More precisely, the only
4--dimensional {\it regular} black hole configurations are those
preserving 4 supersymmetry charges, irrespectively of their higher
dimensional origin. One of the results of \cite{mp1} was  to show that
the most general $1/8$ black hole solution of $N=8$ supergravity can
be related, through a $U$--duality transformation,to a solution of a
suitable  consistent truncation $N=8 \rightarrow N=2$ of the
supergravity theory (corresponding to a Calabi-Yau compactification).
$U$--duality in supergravity denotes the largest global symmetry of
the field equations and Bianchi identities.  We shall recall the main
facts about $U$--duality  orbits of BPS black holes in section 2 and
it will be shown that the problem of studying the most  general $1/8$
BPS black hole in the $N=8$ theory can be reduced to that of finding
the so called  {\it generating solution}, that is  the most general
one modulo $U$--duality transformations, which is a solution  of the
simpler $N=2$ truncation. Once this  solution is found, acting on it
with the maximal compact subgroup $H=SU(8)$  of the $U$--duality group
one generates the general charged  black hole and then acting with the
whole $U$--duality group $U=E_{7(7)}$ one generates the most  general
solution, namely that with fully general asymptotic values of the
scalar fields. In order to define  the action of the $N=8$
$U$--duality group on a solution of the aforementioned $N=2$
truncation, the embedding of the latter  in the $N=8$ theory has to be
defined in a precise group theoretical fashion. This was done in
\cite{mp1} and \cite{bft} using  the solvable Lie algebra  formalism
\cite{mario}. The $N=2$ truncation mentioned above is called the $STU$
model (first studied in \cite{kal1}) and essentially is a $N=2$ supergravity 
coupled to 3 vector multiplets interacting through the Special K\"ahler 
manifold $[SL(2,\IR)/SO(2)]^3$. It has therefore $6$ real scalars 
(3 dilaton--like and 3 axion--like) and four vector fields.  A generic
static, spherically symmetric black hole solution of this model is
then characterized by $4$  electric ($q_\Lambda$) and $4$ magnetic
($p^\Lambda$) quantized charges and a point $\phi_\infty$ in the
scalar manifold  representing the asymptotic behavior of the scalar
fields at infinite radial distance from the center of the
solution. The  generating solution is obtained by fixing the
$[SL(2,\IR)]^3$ duality action on a generic BPS black hole solution,
which is  achieved by first choosing a particular point $\phi_\infty$
of the scalar manifold and then fixing the action of the isotropy
group  $[SO(2)]^3$ on the solution, which amounts to imposing three
suitable conditions on the quantized  charges. These conditions in
turn allow for three conditions on the evolving scalar fields of the
solution, compatible with the field  equations and the BPS
requirement. The generating solution therefore  will be described by
$5$ charges and three independent scalars.  The mathematical framework
in which to construct such a solution has been defined in \cite{bft},
however an explicit solution was  given only in a simplified case
where the 3 dilatons $b_i$ and the 3 axions $a_i$ were separately set
equal to a  single dilaton field $b$ and a single axion field
$a$. This solution, if properly gauge--fixed, would depend only on $2$
parameters.  The aim of the present paper is to relax these
restrictions and compute  the  $5$--parameter generating solution
within the $STU$ model. Other  regular solutions of the same theory,
different from the generating one,  have been considered in other
papers \cite{sabra1}.  
 
As pointed out in  \cite{mp1} and \cite{hull}, the generating solution
represents a pure NS--NS black hole, that is a  black hole whose 10
dimensional microscopic configuration is made only of NS
branes. Indeed, the  generating solution for black holes of the
toroidally compactified type II string theory is the  same as the one
of the toroidally compactified heterotic string and it was already
constructed  in \cite{stu+}. Our point is that, as noticed in
\cite{bala}, as far as the microscopic entropy  counting is concerned
it is difficult to count the states of a pure NS--NS
configuration. Therefore  it is essential to be able to reproduce R--R
configurations which are those represented in terms of  intersecting
D--branes only. This in principle would not be a problem since
$U$--duality does not  make any difference between R--R and NS--NS
fields and one should be able to move from a solution  to any other
acting via $U$--duality transformations. However, as previously
pointed out, in order for this to be possible  one has  not only to
build up the generating solution itself but also to know how the
latter is embedded in the full theory. 

\vskip 5pt
The paper is organized as follows: 

In section 2 we describe the structure of the $U$--duality orbits for $1/8$ BPS black holes 
in terms of $5$ independent invariants and 
characterize the generating solution as a solution depending on $5$ independent
charge parameters, in terms of which the $5$ invariants can be expressed as  
independent functions on the chosen point of the moduli space at infinity.

In section 3 we explicitly construct the 5 parameters generating
solution by solving both the first and second order differential
equations, after performing the aforementioned $3$ parameter gauge
fixing. The need to check that a particular solution of the first
order equations representing the BPS condition fulfill  the second order
field equations as well is due to a known feature of solitonic
solutions in  supergravity that the former equations in general don't
imply the latter.\par Most of the technical aspects concerning both
the derivation of the  first and second order differential equations
(the former representing the BPS conditions, the  latter being the
equations of motion) and the proper formalism needed in order to embed
the $STU$ model solution in the $N=8$ theory properly, have already
been given in \cite{mp1} and \cite{bft}. Hence we shall  skip most
technical details and make use of the results already obtained in
\cite{bft}.

We end in section 4 with few comments and some concluding remarks.


\section{$U$--duality orbits and the generating solution}

Let us recall the main facts about the $U$--duality orbits of 1/8 BPS black holes in the $N=8$
classical supergravity theory \cite{cj},\cite{huto1}. It is well known that the equations of motion 
and the Bianchi identities of the $N=8$ classical supergravity theory in 4--dimensions are invariant 
with respect to the $U$--duality group $E_{7(7)}$. This invariance requires the group $E_{7(7)}$ to 
act simultaneously on both the 70 scalar fields $\phi^{\alpha}$ spanning the manifold 
${\cal M}_{scal}=E_{7(7)}/SU(8)$ and on the vector $\vec Q$ consisting of the 28 electric and 28 
magnetic quantized charges. The $U$--duality group acts on the scalar fields as the isometry group 
of ${\cal M}_{scal} $ and on $\vec Q$ in the ${\bf 56}$ (symplectic) representation. A static, 
spherically symmetric BPS black hole solution is characterized  in general by the vector $\vec{Q}$ 
and a particular point $\phi_\infty$ on the moduli space of the theory whose $70$ coordinates 
$\phi^\alpha_\infty$ are the values of the scalar fields at infinity ($r\rightarrow \infty$). Acting 
on a black hole solution ($\phi_\infty\,,\,\vec Q$) by means of a $U$--duality transformation $g$ one 
generates a new black hole solution ($\phi_\infty^g\,,\,{\vec Q}^g$):
\begin{eqnarray}
\forall g\in G \quad \cases{\phi_\infty\rightarrow \phi_\infty^g(\phi_0)\cr 
\vec{Q}\rightarrow \vec{Q}^g =
{\cal S}(g)\cdot \vec{Q}} \qquad \mbox{where} \qquad {\cal S}(g)\in Sp(56,\IR)
\label{uduality}
\end{eqnarray}
The BPS black hole solutions fill therefore $U$--duality orbits. 

As far as the $1/8$ BPS black holes are concerned these orbits turn out to be parameterized by $5$ 
functions ${\cal I}(\phi_\infty,\vec{Q})_I$ ($I=1,\dots, 5$) which are invariant under the
duality transformations in (\ref{uduality}). These invariants are expressed in terms of the 
$8\times 8$ anti--symmetric central charge matrix $Z_{AB}(\phi_\infty,\vec Q)$ (the antisymmetric 
couple $(AB)$, as $A$ and $B$ run from 1 to 8, labels the representation ${\bf 28}$ of $SU(8)$)  
in the following way (\cite{hull}):
\begin{eqnarray}
{\cal I}_k\, &=&\, {\rm Tr}\left(\bar{Z}Z \right)^k\,\,\,\,\,k=1,\dots,4\nonumber\\
{\cal I}_5\, &=&\, {\rm Tr}\left(\bar{Z}Z\right)^2-\frac{1}{4} ({\rm Tr}\bar{Z}Z)^2+
\frac{1}{96}\left(\epsilon_{ABCDEFGH}Z^{AB}Z^{CD}Z^{EF}Z^{GH}+c.c.\right)\nonumber\\
\label{inv5}
\end{eqnarray}
where $\bar{Z}Z$ denotes the matrix $Z^{AC}Z_{CB}$ and the convention
$Z^{AB}=(Z_{AB})^\star$ is  adopted. Among the ${\cal
I}(\phi_\infty,\vec{Q})_I$ a particular role is played by the  {\it
moduli--independent} invariant ${\cal I}_5(\vec{Q})$ that is the {\it
quartic invariant} (which will be denoted in the  sequel also by
$P_{(4)}(\vec{Q})$ in order to refer to its group theoretical meaning
)  of $E_{7(7)}$ whose value is related to the entropy of the black
hole \cite{fer},\cite{entropy}. As pointed out in  \cite{maldasergio},
$P_{(4)}$ must be non--negative in order for the solution to be BPS
($P_{(4)}\ge 0$).  For a fixed value  of ${\cal I}_5(\vec{Q})$ the
inequivalent orbits are parameterized by  the remaining four invariants
${\cal I}(\phi_\infty,\vec{Q})_k$, ($k=1,\dots, 4$). The behavior of
the scalars describing the regular solutions with fixed entropy is
schematically represented in Figure $1$ in which the  scalar fields
flow from their boundary values $\phi_\infty$ at infinity which span
${\cal M}_{scal}$  (the disk) to their fixed values $\phi_{fix}$ at
the horizon $r=0$ \cite{fer}. It should be  understood, of course,
that the $\phi$ axis is a $n$--dimensional space, where $n$ is the
dimension  of ${\cal M}_{scal}$. The invariants ${\cal
I}(\phi_\infty,\vec{Q})_I$ turn out to be independent  functions of
the quantized charges in any generic point $\phi_\infty$ of the
moduli--space except  for some ``singular'' points where the number of
truly independent invariants could be less than  five. This is the
case, for example, of the point $\phi_{\infty}=\phi_{fix}$
parameterized by the  fixed values of the scalar fields at the
horizon. In this point the {\it only} independent invariant  is the
moduli--independent one, ${\cal I}_5$. We will come back on this at
the end of this section.
\begin{figure}
\vskip 20pt
\input epsf
\epsfxsize=250pt
\centerline{\epsffile{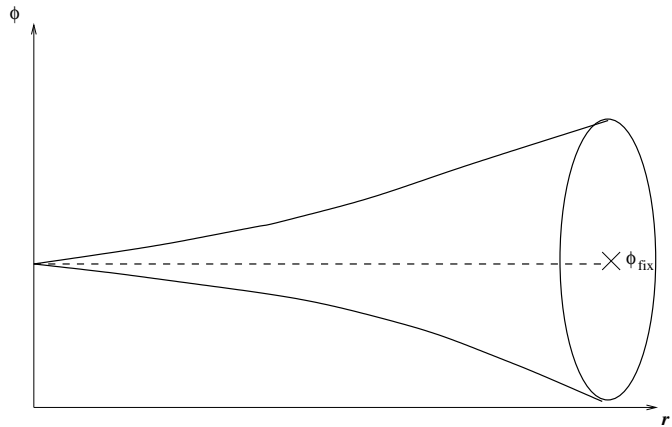}}
\caption{\scriptsize the radial dependence of the scalar fields from the horizon to the 
spatial infinity.}
\end{figure}

The generating solution may be characterized as the solution depending on the the minimal number 
of parameters sufficient to obtain all possible $5$--plets of values for the $5$ invariants on a 
particular point $\phi_\infty\neq \phi_{fix}$ of the moduli--space (a possible vacuum of the theory).
From the above characterization it follows that the whole
$U$--duality orbits of $1/8$ BPS black hole solutions may be
constructed by acting by means of $E_{7(7)}$ transformations
(\ref{uduality}) on the generating one. In particular, if we
focus on $1/8$ BPS black hole solutions having a fixed value of the
entropy (proportional to the square root of ${\cal I}(\vec Q)_5$) and
on a particular bosonic vacuum for the theory specified by a point $\phi_\infty$
in the moduli--space, by acting only on the charges of the generating
solution with the $U$--duality group, it would be possible to
construct the whole spectrum of $1/8$ BPS solutions of the theory
realized in the chosen vacuum $\phi_\infty$ (see Figure 2). Since in a particular point 
$\phi_\infty \neq \phi_{fix}$ on ${\cal M}_{scal}$ the minimum number of parameters a solution 
should depend on in order to reproduce all the $5$--plets of values for the independent invariants 
is obviously five, we expect  the charge vector $\vec Q$ of the generating solution 
$(\phi_\infty ,\vec Q)$ to depend on five independent charges. 
\begin{figure}
\vskip 20pt
\input epsf
\epsfxsize=250pt
\centerline{\epsffile{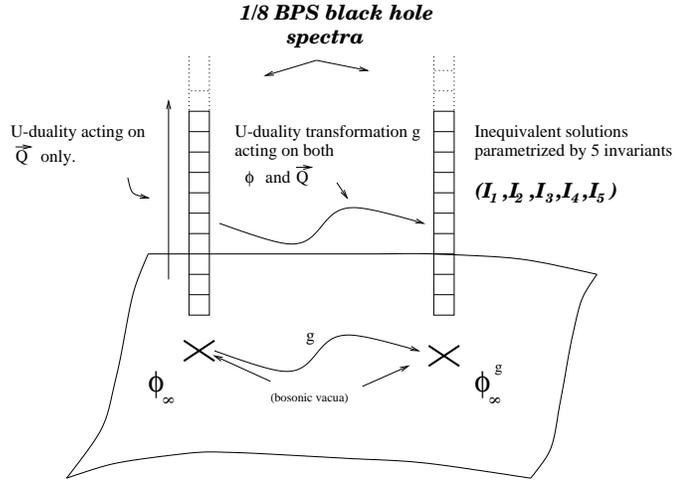}}
\caption{\scriptsize the scalar moduli--space ${\cal M}_{scal}$ and the action of the $U$--duality 
group on a generic point $\phi_{\infty}$. The tower on each point of ${\cal M}_{scal}$ represents 
the $1/8$ BPS black hole spectrum of the theory realized on that bosonic background. In a generic 
point $\phi_{\infty}$ it describes a ``snapshot'' of the $U$--duality orbit
of this particular kind of solutions. Acting on the charges only, one generates all black holes 
having the same values of the asymptotic fields and moves in the tower at a given $\phi_{\infty}$. 
Acting on both the charges and the moduli one moves in ${\cal M}_{scal}$ but the solution does not 
change its ADM mass (this being a $U$--duality invariant quantity). Since the $U$--duality orbit is 
characterized by $5$ invariants ${\cal I}_I$, in a point $\phi_{\infty}$ in which ${\cal I}_I$ are 
independent functions of the charges $\vec{Q}$ 
the generating solution may be characterized as a {\it minimal} set of solutions in the corresponding 
tower on which the five invariants assume all possible values (compatible with the BPS condition). 
Therefore the generating solution should depend only on five charge parameters and, by acting 
just on the latter by means of the $U$--duality group (``vertical action'' in the figure), one is able 
to reconstruct {\it all} the sates of the tower.}
\end{figure}

In order to motivate, in brief, the result obtained in \cite{mp1} according to which  the
generating solution for $1/8$ BPS black holes in the $N=8$ theory is
described within a suitable $N=2$ truncation of the theory (the $STU$
model), let us notice that the $5$ quantities in eq. (\ref{inv5}) are
invariant with respect to the action of $SU(8)$ on $Z_{AB}$.
In particular, by means of a $48$--parameter $SU(8)$ transformation,
the central charge matrix can be brought to its {\it normal form} in which it is skew--symmetrized 
with complex eigenvalues $( Z_k)$, $k=1,\dots,4$ ($|Z_4|>|Z_3|\ge |Z_2|\ge |Z_1|$). As a consequence 
of this rotation (see \cite{mp1} and \cite{bft} for a detailed discussion) the central charge 
eigenvalues end up depending on just $6$ (dynamical) scalar fields and $8$ quantized charges which 
characterize the solutions of an $STU$ model suitably embedded in the original theory. 
Within this truncation, the $6$ scalar fields ($3$ dilatons $b_i$ and $3$ axions $a_i$) 
belong to $3$ vector multiplets and span a manifold ${\cal M}_{STU}=[SL(2,\IR)/SO(2)]^3$, while 
the $4$ electric charges $q_\Lambda$ and $4$ magnetic charges $p^\Lambda$ ($\Lambda=0,\dots,3$) 
transform in the ${\bf (2,2,2)}$  of $[SL(2,\IR)]^3$. In the framework of the $STU$ model, the 
central charge eigenvalues $Z_4(a_i,b_i,\vec Q)$ and $Z_i(a_i,b_i,\vec Q)$ ($i=1,2,3$) are, 
respectively, the local realization on moduli space ${\cal M}_{STU}$ of the $N=2$ supersymmetry 
algebra central charge $Z$ and of the $3$ {\it matter} central charges associated with the $3$ 
matter vector fields (which are related to the central charge via the following relation:
$Z^i(z,\bar{z},{p},{q})=h^{ij^\star}\nabla_{j^\star}\bar{Z}(z,\bar{z},{p},{q})$, $h_{ij^\star}$ 
being the K\"ahler metric on ${\cal M}_{STU}$ and $z_i=a_i+ib_i$). 
On the $1/8$ BPS black hole solutions these four eigenvalues are in general independent in a generic 
point of the moduli--space and the BPS condition reads:
\begin{equation}
M_{ADM}=\lim_{r\rightarrow \infty}|Z_4(a_i,b_i,\vec Q)|
\end{equation}
Since the $SU(8)$ transformation used to define the $STU$ truncation of
the original theory did not affect the values of the $5$ invariants
in eq. (\ref{inv5}), the latter are expected to assume all possible $5$--plets of
values on BPS solutions of this theory. From this we conclude that {\it the generating solution 
for $1/8$ BPS black holes in the $N=8$ theory is a solution of the $STU$ truncation as well.}

In the framework of the $STU$ model the five invariants in eq. (\ref{inv5})
are rewritten in the following form:
\begin{eqnarray}
{\cal I}(\phi_\infty,\vec{Q})_k\,&=&\,\sum_{k^\prime=1}^{4}\vert Z_{k^\prime}\vert^{2k}\nonumber\\
{\cal I}(\vec{Q})_5\, &=&\,\sum_{k=1}^{4}\vert Z_k\vert^4-2\sum_{k_1>k_2=1}^{4}\vert
Z_{k_1}\vert^2\vert Z_{k_2}\vert^2+ 4\left( Z_1^\star Z_2^\star Z_3^\star Z_4^\star+ 
Z_1 Z_2 Z_3 Z_4\right)
\label{invstu}
\end{eqnarray}
However in this model there is still a residual invariance of the
above quantities represented by the $3$ parameter group $[SO(2)]^3$,
isotropy group of the scalar  manifold ${\cal M}_{STU}$ and subgroup
of $SU(8)$. It acts on the four phases $\theta_k$ of the  central
charge eigenvalues $Z_k$ leaving the overall phase $\theta=\sum_k
\theta_k$ invariant. The generating solution is obtained by fixing
this gauge freedom and therefore it depends, consistently with what
stated above, on $5$ parameters represented  by the four norms of the
central charge eigenvalues $\vert Z_{k}\vert$ plus the overall phase
$\theta$. These quantities are $U$--duality invariants as well.  It
can be shown indeed that the norms $\vert Z_{k}\vert$ may be expressed
in terms of the four invariants ${\cal I}_k$ ($k=1,2,3,4$) while the
overall phase is contained in the expression of the Pfaffian in ${\cal
I}_5$ and thus is an invariant quantity as well which is expressed in
terms of all the five ${\cal I}_I$. Indeed, see eq.s (\ref{inv5}) and
(\ref{invstu}):
\begin{eqnarray}
\frac{1}{96}\left(\epsilon_{ABCDEFGH}Z^{AB}Z^{CD}Z^{EF}Z^{GH}+c.c.\right)\,  &=&\, 
4\left( Z_1^\star Z_2^\star Z_3^\star Z_4^\star+ Z_1 Z_2 Z_3 Z_4\right)\,=\nonumber\\
&& 2\vert Z_1 Z_2 Z_3 Z_4 \vert {\rm cos}\theta
\label{pf}
\end{eqnarray}
The moduli independent invariant ${\cal I}_5$, computed in the $STU$ model, is the quartic invariant 
of the ${\bf (2,2,2)}$ of $[SL(2,\IR)]^3$ and it is useful to express it in a form which is intrinsic 
to this representation. We may indeed represent the vector $\vec{Q}=(p^\Lambda,q_\Sigma)$ as a tensor
$q^{\alpha_1\alpha_2\alpha_3}$ where $\alpha_i=1,2$ are the indices of the
${\bf 2}$ of each $SL(2,\IR)$ factor. The invariants are constructed
by contracting the indices of an even number $2m$ of $q^{\alpha_1\alpha_2\alpha_3}$ with $3m$ 
invariant matrices $\epsilon_{\alpha_i\beta_i}$.
This contraction gives zero for $m$ odd while for $m$ even one finds:
\begin{eqnarray}
P_{(2)}(p,q)\,&=&\, q^{\alpha_1\alpha_2\alpha_3}q^{\beta_1\beta_2\beta_3}
\epsilon_{\alpha_1\beta_1}\epsilon_{\alpha_2\beta_2}\epsilon_{\alpha_3\beta_3}\,=\,0\nonumber\\
P_{(4)}(p,q)\,&=&\, q^{\alpha_1\alpha_2\alpha_3}q^{\beta_1\beta_2\beta_3}
q^{\gamma_1\gamma_2\gamma_3}q^{\delta_1\delta_2\delta_3}
\epsilon_{\alpha_1\beta_1}\epsilon_{\alpha_2\beta_2}\epsilon_{\gamma_1\delta_1}
\epsilon_{\gamma_2\delta_2}\epsilon_{\alpha_3\gamma_3}
\epsilon_{\beta_3\delta_3}\,=\nonumber\\
&&4(p^3q_0+q_1q_2)(p^1p^2-p^0q_3)-(p^0q_0+p^1q_1+p^2q_2-p^3q_3)^2\nonumber\\
P_{(8)}(p,q)\,&=&\,c\times (P_4(p,q))^2\nonumber\\
P_{(12)}(p,q)\,&=&\,c^\prime \times (P_4(p,q))^3\nonumber\\
&&\dots
\label{invariants}
\end{eqnarray}
It can be shown rigorously that the quartic invariant written above is the {\it only} independent 
invariant of this representation, that is any other invariant may be expressed as powers of it. The 
square root of $P_{(4)}$ is proportional to the entropy of the solution and its expression is 
consistent  with the result of \cite{kal1}. In terms of the $8$ quantized charges the components of 
$q^{\alpha_1\alpha_2\alpha_3}$ are:
\begin{eqnarray}
q^{1,1,1}\,&=&\,p^0\,;\, q^{2,1,1}\,=\,p^1\,;\,q^{1,2,1}\,=\,p^2\,;\,
q^{1,1,2}\,=\,p^3\,;\nonumber\\
q^{2,2,2}\,&=&\,-q_0\,;\, q^{1,2,2}\,=\,q_1\,;\,q^{2,1,2}\,=\,q_2\,;\,
q^{2,2,1}\,=\,q_3
\end{eqnarray}

According to the above characterization of the generating solution, it
is apparent therefore that the black hole found in \cite{bft}
represents just a particular (regular)  solution characterized by just
$2$ invariant parameters ($|Z_1|=|Z_2|=|Z_3|$ and $|Z_4|$),  although
depending on three charges. That solution was obtained by setting
$S=T=U$ and  $p^1=p^2=p^3=p$, $q_1=q_2=q_3=q$, $p^0=0$. Therefore,
acting on it by means of an $U$--duality  transformation, it is clear
from our previous discussion that it would be possible to span only  a
$2$--parameter sub--orbit of the whole $U$--duality orbit. The same
argument holds for all the  regular $1/8$ BPS black holes
characterized by  less than $5$ independent invariant parameters,   in
particular for the double fixed solution found in \cite{kal1}. This is
in fact a very particular  solution in which the scalar fields do not evolve 
and their values at infinity coincide with their fixed ones at the horizon
$\phi_{\infty}=\phi_{fix}$. At this point the matter central charges
vanish  ($Z_i(\phi_{fix},\vec Q)=0$, $i=1,2,3$) and the central charge
$Z_4(\phi_{fix},\vec Q)$ becomes  proportional to ${\cal I}(\vec
Q)_5^{1/4}$ which represents therefore the only invariant  parameter
characterizing the spectrum of $1/8$ BPS solutions on this particular
point of the  moduli--space. The $5$--plet of invariants $({\cal
I}_1,{\cal I}_2,{\cal I}_3,{\cal I}_4,{\cal I}_5)$  on this solution
have the specific simple form $(a,a^2,a^3,a^4,a^2)$, $a$ being a
constant depending on the $8$  charges and proportional to the entropy.

Summarizing, the three main facts reviewed in the present section are:
\begin{itemize}
\item{The $U$--duality orbits of $1/8$ BPS black hole solutions are characterized by five $U$--duality 
invariants ${\cal I}_I$.}
\item{ The generating solution $(\phi_{\infty},\vec{Q})$ is characterized by a generic point on the 
moduli space at infinity $\phi_{\infty}\neq\phi_{fix}$ (in which the five invariants are actually 
independent functions of the charges $\vec{Q}$) and by five independent charge parameters $\vec{Q}$ 
such that, by 
varying the latter one obtains all possible combinations of values of ${\cal I}_I$ consistent
with the BPS condition (${\cal I}_5\ge 0$).} 
\item{The generating solution of 1/8 susy preserving black holes of $N=8$ supergravity can be 
thought of as a 1/2 preserving solution of an $STU$ model suitably embedded in
the original theory \cite{bft}. In this framework, which is that we will deal with, the five 
invariants ${\cal I}_I$ can be expressed as proper combinations of the norms of the four central 
charges $|Z_k|$ (the supersymmetry and the three matter ones corresponding to $k=4$ and $k=1,2,3$ 
respectively) and their overall phase $\theta$, according to eqs. (\ref{invstu}) and (\ref{pf}).}
\end{itemize}

\section{The 5 parameter generating solution}

In the present section we shall compute the $5$--parameter generating solution
as a $1/2$ BPS black hole solution of the $STU$ model.

The property a BPS saturated state has of preserving a fraction of the
original supersymmetries can be characterized, on a bosonic background, by requiring that 
the supersymmetry transformations of all the fermion fields vanish along a suitable direction in 
the supersymmetry parameter space (Killing spinor):
\begin{eqnarray}
\delta_{\epsilon}\mbox{fermions}\, &=&\, 0\nonumber\\
\gamma^0 \,\epsilon_{A} \,& =&\,  \pm \mbox{i}\, \epsilon_{AB}
\,  \epsilon^{B} \quad \mbox{if} \quad A,B=1,2
\label{killing}
\end{eqnarray}
The first of these equations is equivalent to a system of first order differential equations on 
the background fields. In particular the system of equations in the scalar fields has a fixed point, 
which will be denoted by $\phi_{fix}=(b_i^{fix},a_i^{fix})$, towards  which the solution
 $\phi(r)=(b_i(r),a_i(r))$ flows to at the horizon $r\rightarrow 0$ \cite{fer}.

The way eq. (\ref{killing}) has been written expresses the reality condition for $Z_4(\phi, p,q)$ and 
it amounts to fix one of the three $SO(2)$ gauge symmetries of $H$ already giving therefore a 
condition on the $8$ charges and the scalar fields. In general however, as pointed out in 
\cite{moore}, one should in fact consider a more general form for the killing spinor condition 
\footnote{This result was eventually used in \cite{feco} and \cite{sabra2}.}, namely:
\begin{equation}
\gamma^0 \,\epsilon_{A} = \pm \mbox{i}\,\frac{Z_4}{|Z_4|}\, \epsilon_{AB}\,  \epsilon^{B}
\end{equation}
and, as noticed by G. Moore, imposing the reality of the central charge could in principle imply 
that some topologically non--trivial solutions are disregarded. Nevertheless studying such a special 
class of solutions is not among the purposes of our present investigation and therefore we shall 
choose the supersymmetry central charge to be real (${\rm Im}(Z_4)=0$). 

Let us consider the gauge fixing procedure in detail. The four central charges 
$Z_k(\phi,\vec{Q})$ of the $STU$ model, depending on the asymptotic values of the six 
scalars $\phi_{\infty}=(a^{\infty}_i,b^{\infty}_i)$ and $8$ charges $\vec{Q}=(p^\Lambda,q_\Sigma)$, 
transform under $[SL(2,\IR)]^3$ duality (\ref{uduality}) as follows:
\begin{eqnarray}
\forall g\in \left[SL(2,\IR)\right]^3&&Z_k(\phi^g,\vec{Q}^g)\,=\,h_g\cdot Z_k(\phi,\vec{Q})\nonumber\\
h_g\in SO(2)^3 && h_g\cdot Z_k\equiv e^{{\rm i}\delta_k^g}Z_k
\label{phases}
\end{eqnarray}
Hence  an $[SL(2,\IR)]^3$ duality transformation on the moduli at
infinity and on the quantized charges  amounts to an $[SO(2)]^3$ phase
transformation on the four central charges. This holds true in
particular if we consider $g\in [SO(2)]^3$. It follows that the
$[SO(2)]^3$ gauge fixing may be  achieved by either imposing three
suitable conditions on the phases of the central charges, or
alternatively fixing the $[SO(2)]^3$ action on $\vec{Q}$ on a chosen
point $\phi^0_{\infty}$ of  the moduli space at infinity. We shall
pursue the latter way which amounts to impose three suitable
conditions on the quantized charges. These conditions will not be
derived with group theoretical arguments; it will suffice, as it has
been pointed out in the last section, to show that  the five
invariants, computed in $\phi_{\infty}$,  are independent functions of
the remaining five charges. 

Since, on one hand, the two--fold action of a duality transformation
(and in particular of a $[SO(2)]^3$ transformation) on both the
quantized charges and the scalar fields is an invariance of the
equations of motion, and on the other hand the charges $\vec{Q}$ are
``constants of motion'' (with respect to the $r$--evolution), we
expect that {\it the three gauge  fixing conditions on the electric
and magnetic charges have a counterpart in three $r$--independent
conditions on the fields $\phi(r)$, such that the restricted system of
scalar fields and vector  fields is still a solution of the field
equations.}   After the gauge fixing procedure therefore we would
expect the solution to be described  by five independent charges,
three real scalar fields and the metric function ${\cal U}(r)$.  The
evolution of the  latter four fields will be described in terms of
four  different harmonic functions.

Let us now recall the main positions on the background fields used to derive the first order equations 
from equation (\ref{killing}).

The first of eqs. (\ref{killing}) can be specialized to the supersymmetry transformations of the 
gravitino and gaugino within the $STU$ model in the following way:
\begin{eqnarray}
\delta_{\epsilon} \psi_{A\vert \mu}\,&=&\,\nabla_{\mu}\epsilon_A-\frac{1}{4}T^-_{\rho \sigma}
\gamma^{\rho \sigma}\gamma_{\mu}\epsilon_{AB}\epsilon^B\,=\, 0\nonumber\\
\delta_{\epsilon}\lambda^{i\vert A}\,&=&\,{\rm i}\nabla_{\mu}z^i\gamma_{\mu}\epsilon_A+
G^{-\vert i}_{\rho \sigma}\gamma^{\rho \sigma}\epsilon^{AB}\epsilon_B\,=\, 0
\label{newkill}
\end{eqnarray}
where $i=1,2,3$  labels the three matter vector fields, $A,B=1,2$ are the $SU(2)$ R-symmetry indices
and $T^-_{\rho \sigma}$ and $G^{-\vert i}_{\rho \sigma}$ are the graviphoton and matter field strengths
respectively (the $-$ sign stands for the anti--self dual part). 
According to the procedure defined in
\cite{bft},  we adopt the following ans\"atze for the
vector fields:
\begin{eqnarray}
\label{strenghtsans}
F^{-\vert \Lambda}\,&=&\, \frac{t^\Lambda (r)}{4\pi}E^- \;,\,
t^\Lambda(r)\,=\, 2\pi(p^\Lambda+{\rm i}\ell ^\Lambda (r))\nonumber\\
F^{\Lambda}\,&=&\,2{\rm Re}F^{-\vert \Lambda}\,\,;\,\,
\tilde{F}^{\Lambda}\,=\,-2{\rm Im}F^{-\vert \Lambda}\nonumber\\
F^{\Lambda}\,&=&\,\frac{p^{\Lambda}}{2r^3}\epsilon_{abc}
x^a dx^b\wedge dx^c-\frac{\ell^{\Lambda}(r)}{r^3}e^{2\cal {U}}dt
\wedge \vec{x}\cdot d\vec{x}\nonumber\\
\tilde{F}^{\Lambda}\,&=&\,-\frac{\ell^{\Lambda}(r)}{2r^3}\epsilon_{abc}
x^a dx^b\wedge dx^c-\frac{p^{\Lambda}}{r^3}e^{2\cal {U}}dt\wedge \vec{x}\cdot d\vec{x}
\end{eqnarray}
where
\begin{eqnarray}
E^-\,&=&\,\frac{1}{2r^3}\epsilon_{abc}
x^a dx^b\wedge dx^c+\frac{{\rm i}e^{2\cal {U}}}{r^3}dt\wedge \vec{x}\cdot d\vec{x}\,=\,\nonumber\\
&&E^-_{bc}dx^b\wedge dx^c+2E^-_{0a}dt\wedge dx^a \nonumber\\
4\pi\,&=&\,\int_{S^2_\infty}E^-_{ab} dx^a\wedge dx^b \nonumber\\
\end{eqnarray}
The moduli--independent quantized charges  $(p^\Lambda,q_\Sigma)$ and the moduli--dependent electric charges $\ell_\Sigma (r)$ \cite{bft} are obtained by the following integrations: 
\begin{eqnarray}
4\pi p^\Lambda\,&=&\,\int_{S^2_r}F^{\Lambda}\, = \,
  \int_{S^2_\infty}F^{\Lambda}\,=\,2{\rm Re}t^\Lambda\nonumber\\
4\pi q_\Sigma\,&=&\,\int_{S^2_r}G_{\Sigma}\, = \,
  \int_{S^2_\infty}G_{\Sigma}\nonumber\\
4\pi \ell^\Lambda (r)\,&=&\,-\int_{S^2_r}\tilde{F}^{\Lambda}\,=\,2
{\rm Im}t^\Lambda
\label{cudierre}
\end{eqnarray}
where $S^2_r$ and $S^2_\infty$ denote the spheres centered in $r=0$ of radius $r$ and $\infty$ 
respectively. The expression of the moduli--dependent charges $\ell_\Sigma (r)$ in terms of the 
scalars $(a_i,b_i)$ and the charges $(p^\Lambda,q_\Sigma)$ is given in the appendix.

As far as the metric $g_{\mu \nu}$, the scalars $z^i=a_i+{\rm i}b_i$, parameterizing 
$[SL(2,\IR)/SO(2)]^3$, and the Killing spinors $\epsilon_A (r)$
are concerned, the ans\"atze we adopt are the following:
\begin{eqnarray}
ds^2\,&=&\,e^{2{\cal U}\left(r\right)}dt^2-e^{-2{\cal U}\left(r\right)}d\vec{x}^2 ~~~~~
\left(r^2=\vec{x}^2\right)\nonumber\\
z^i\,&\equiv &\, z^i(r)\nonumber\\
\epsilon_A (r)\,&=&\,e^{f(r)}\xi_A~~~~~~~~~\xi_A=\mbox{constant}\nonumber\\
\gamma_0 \xi_A\,&=&\,\pm {\rm i}\epsilon_{AB}\xi^B
\label{fishlicense}
\end{eqnarray}
 Substituting the above ans\"atze in eqs. (\ref{newkill}), after some algebra, 
one obtains an equivalent system of first order differential equations on the background fields 
of the form:
\begin{eqnarray}
\frac{dz^i}{dr}\, &=&\,
\mp \left(\frac{e^{{\cal U}(r)}}{4\pi r^2}\right) h^{ij^\star}\nabla_{j^\star}
\bar{Z_4}(z,\bar{z},{p},{q})
\nonumber\\
\frac{d{\cal U}}{dr}\, &=&\, \mp \left(\frac{e^{{\cal U}(r)}}{r^2}\right)|Z_4(z,\bar{z},{p},{q})|
\label{eqs122}
\end{eqnarray}
where the supersymmetry central charge $Z_4$ has the following expression: 
\begin{eqnarray}
Z_4(z,\bar{z},{p},{q})\,=\,-\frac{1}{4\pi}\int_{S^2}T^-\,=
\,M_\Sigma {p}^\Sigma-L^\Lambda {q}_\Lambda
\nonumber
\end{eqnarray}
The vector $(L^\Lambda(z,\bar{z}),M_\Sigma(z,\bar{z}))$ is the covariantly holomorphic section 
on the symplectic bundle defined on the Special K\"ahler manifold ${\cal M}_{STU}$.

As already stressed, in order to find a proper solution we need also the equations of motion that 
must
be satisfied together with the first order ones. The former can be derived from an $N=2$ pure
supergravity action coupled to 3 vector multiplets (see \cite{N=2} for notation):
\begin{eqnarray}
\label{action}
S\,&=&\, \int d^4x\sqrt{-g}\,{\cal L}\qquad\quad \mbox{where}\nonumber\\
{\cal L}\,&=&\,R[g]+h_{ij^\star}(z,\bar z)\partial_\mu z^i \partial^\mu
\bar{z}^{j^\star}
+\left({\rm Im}{\cal N}_{\Lambda\Sigma}F^{\Lambda}_{\cdot\cdot}
F^{\Sigma\vert\cdot\cdot}+{\rm Re}
{\cal N}_{\Lambda\Sigma}F^{\Lambda}_{\cdot\cdot}\tilde{F}^{\Sigma\vert\cdot\cdot}\right)
\end{eqnarray}
It was shown in \cite{bft} that the Maxwell equations are automatically
satisfied by the ans\"atze (\ref{strenghtsans}).
What really matters then are the scalar and Einstein equations which should be fulfilled by 
our ans\"atze.  Their explicit form have been computed in \cite{bft} and we report them in 
the appendix.

\subsection{Gauge fixing conditions and invariants}
The generating solution we are going to define is characterized by the 
following  conditions on the quantized charges and on the scalar fields:\\
\underline{ On the charges:}
\begin{eqnarray}
p^0\,=\,0\;&,&\;\frac{p^1}{p^3}=\frac{ p^3}{p^2}\,\equiv \,\alpha\;\;,\;\;
p^1 q_1 + p^2 q_2 +p^3 q_3 \,=\,0 \quad \mbox{with}\;\;\alpha\, \neq\, 0\nonumber\\
\label{charcond}
\end{eqnarray}
\underline{ On the fields:}
\begin{eqnarray}
b_1\,=\, \alpha b\;\;,\;\;b_2\,&=&\,\frac{ b}{\alpha}\;\;,\;\;b_3\,\equiv \,b
\label{scalcond1}
\end{eqnarray}
\begin{eqnarray}
\frac{a_1}{\alpha}+\alpha a_2 + a_3\,=\, 0\nonumber\\
\label{imz0}
\end{eqnarray}
The relation (\ref{imz0}) on the axions derives from the reality condition on the 
supersymmetry central charge $Z_4$:
\begin{equation}
{\rm Im} Z_4\,=\,0
\end{equation}  
once the conditions (\ref{charcond}) on the charges and the 
 two conditions on the scalars
in eqs (\ref{scalcond1}) are taken into account. \par
One may check that the positions (\ref{charcond}),(\ref{scalcond1}) and 
(\ref{imz0}) are indeed consistent with the field equations and with the 
system of first order equations (\ref{mammamia}). In particular they 
are fulfilled by the fixed point values of the scalar fields $(a_i^{fix},b_i^{fix})$:
\begin{eqnarray}
a^{fix}_1\, &=&\,-\frac{\alpha^2 q_1}{p^3}\,\,;\,\,a^{fix}_2\,=\,\frac{(\alpha q_1+q_3)}
{\alpha p^3}\,\,;\,\,a^{fix}_3\,=\,-\frac{q_3}{p^3}\nonumber\\
b_1^{fix}\, &=&\,\alpha b_{fix}\,\,;\,\,b_2^{fix}\,=\,\frac{b_{fix}}{\alpha}\,\,;\,\,
b_2^{fix}\,=\,b_{fix}\nonumber\\
b_{fix}^2\, &=&\,\frac{1}{4 (p^3)^4}P_{(4)}(p,q)\nonumber\\
P_{(4)}(p,q)\, &=&\,4(p^3)^3\left( q_0-\frac{q_1^2\alpha^2}{p^3}-\frac{q_3}{ p^3}
\left(\alpha q_1 +q_3\right)\right)
\label{fixedv}
\end{eqnarray}
where $P_{(4)}(p,q)$ was defined in eqs. (\ref{invariants}).

A solution consistent with (\ref{charcond}),(\ref{scalcond1}) and 
(\ref{imz0}) will be described by four independent fields, say $a_1,a_2,b,{\cal U}$
and five independent parameters, say $\alpha, p^3,q_0,q_1,q_3$.
As previously pointed out, in order for our solution to be BPS it is necessary that 
$P_{(4)}(p,q)\ge 0$ which provides an inequality condition on the five parameters 
$\alpha,p^3,q_0,q_1,q_3$.
 
To check that conditions  (\ref{charcond}),(\ref{scalcond1}) and 
(\ref{imz0}) actually fix the $[SO(2)]^3$ gauge, let us compute the five 
invariants $|Z_k|, {\rm tg}(\theta)$ in a suitable point of the moduli space at infinity 
$\phi_{\infty}$ which will characterize the asymptotic behavior of our generating solution. 
For simplicity we shall choose:
\begin{equation}
\phi_{\infty}\,=\,\left(a_i^{\infty}=0\,;\,b^{\infty}_1=-\alpha\,,\,b^{\infty}_2=-1/\alpha\,,\,
b^{\infty}_3= b^{\infty}=-1\right)
\label{inphi}
\end{equation}
It is useful to express the invariants in terms of the following quantities:
\begin{eqnarray}
x\, &=&\, {\rm Re}(Z_1)\,=\,\sqrt{-\frac{b}{2}}2p^3 \left( a_1-a_1^{fix}\right)\nonumber\\
y\, &=&\, {\rm Re}(Z_2)\,=\,\sqrt{-\frac{b}{2}}2p^3 \left( a_2-a_2^{fix}\right)\nonumber\\
w\, &=&\, {\rm Im}(Z_3)\,=\,\sqrt{-\frac{1}{2b}}\left( 
\sum_i q_i a_i-\sum_{P(ijk)}p^ia_j a_k +q_0 -b^2 p^3\right)\nonumber\\
v\,&=&\,{\rm Re}(Z_4)\, =\, \sqrt{-\frac{1}{8b^3}}
\left( \sum_i q_i a_i-\sum_{P(ijk)}p^ia_j a_k +q_0 +3b^2 p^3\right)
\label{xywv}
\end{eqnarray}
where $P(ijk)$ denotes the three cyclic permutations of the indices 
$(ijk)$. It is straightforward to show that the five invariants have the following 
expressions in terms of the above quantities:
\begin{eqnarray}
|Z_1|^2\, &=&\, \alpha^2 w^2+x^2\nonumber\\
|Z_2|^2\, &=&\, \frac{w^2}{\alpha^2}+y^2\nonumber\\
|Z_3|^2\, &=&\, w^2+\left(y\alpha+\frac{x}{\alpha}\right)^2\nonumber\\
|Z_4|^2\, &=&\,v^2\nonumber\\
{\rm tg}(\theta)\, &=&\,\frac{w(x^2 +\alpha^4 y^2 +\alpha^2 (w^2+xy)) }{\alpha
x y (x+\alpha^2 y)}
\label{inva}
\end{eqnarray}
where we have used the property that ${\rm Re}(Z_1)$, ${\rm Re}(Z_2)$ and ${\rm Re}(Z_3)$ have to 
fulfill the same linear relation (\ref{imz0}) as the axions $a_i$.
An important feature of the system (\ref{inva}) is to have a real solution in terms of 
$x,y,w,v,\alpha$ for {\it any} $5^{plet}$ of values of the invariants. Particular care, however, 
has to be taken when dealing with the case in which one or more of the matter central charges vanish. 
The corresponding solution of the system (\ref{inva}) is obtained by defining suitable limits of 
the quantities $x,y,w,v,\alpha$ in which it can be shown that the generating solution is still 
regular. The case for which $|Z_i|=0$, $i=1,2,3$ is easily solved by setting $x=y=w\rightarrow 0$, 
and corresponds to the double--fixed solution.

Now let us compute the invariants in (\ref{inva}) on our solution, which is characterized by the 
above defined point $\phi_{\infty}$ in the moduli space, eq. (\ref{inphi}), and by the five 
independent parameters $\alpha,p^3,q_0,q_1,q_3 $ left over after applying the conditions 
(\ref{charcond}) on the $8$ quantized charges. To show that the five invariants in 
$\phi_{\infty}$ are actually independent functions of  $\alpha,p^3,q_0,q_1,q_3 $, we simply 
need to show that the quantities $x,y,w,v,\alpha$ are. Specializing equations (\ref{xywv}) to 
the point $\phi_{\infty}$ in (\ref{inphi}) one obtains:
\begin{eqnarray}
x\,&=&\, \sqrt{2}\alpha^2 q_1 \nonumber\\
y\,&=&\, -\frac{ \sqrt{2}}{\alpha}\left(\alpha q_1+ q_3\right)\nonumber\\
w\,&=&\, \frac{1}{\sqrt{2}}\left(q_0-p^3\right)\nonumber\\
v\,&=&\, w+2\sqrt{2}p^3
\label{ammersa2}
\end{eqnarray}
the above system has clearly a solution in terms of  $\alpha,p^3,q_0,q_1,q_3 $
for any combinations of real values for $\alpha,x,y,w,v$. 

Therefore we have shown that by varying the five parameters  $\alpha,p^3,q_0,q_1,q_3 $, one may 
cover the whole spectrum of values 
for the five invariants in (\ref{inva}) (or equivalently ${\cal I}_I$, $I=1,\dots,5$ defined in 
section $2$) describing all possible $U$--duality inequivalent solutions.   

\subsection{The solution}
Implementing the conditions (\ref{charcond}),(\ref{scalcond1}) and (\ref{imz0}), the system of 
first order equations (\ref{mammamia}) simplifies dramatically and reduces to the following form:
\begin{eqnarray}
\frac{db}{dr}\, &=&\, \left(\frac{e^{\cal U}}{r^2}\right)\frac{1}{\sqrt{-2b}}\left(
F(a_i)-b^2p^3\right)\nonumber\\
\frac{da_1}{dr}\, &=&\, \left(\frac{e^{\cal U}}{r^2}\right)
\sqrt{-b}\sqrt{2}p^3\left(a_1-a_1^{fix}\right)\nonumber\\
\frac{da_2}{dr}\, &=&\, \left(\frac{e^{\cal U}}{r^2}\right)\sqrt{-b}\sqrt{2}p^3
\left(a_2-a_2^{fix}\right)\nonumber\\
\frac{d{\cal U}}{dr}\, &=&\, \left(\frac{e^{\cal U}}{r^2}\right)\frac{1}{2\sqrt{-2b^3}}\left(
F(a_i)+3b^2p^3\right)\nonumber\\
F(a_i)\,&=&\, \sum_i q_i a_i-\sum_{P(ijk)}p^ia_j a_k  +q_0
\label{1eqex}
\end{eqnarray}
To solve the above equations it is useful to start defining the function $h(r)=e^{\cal U}\sqrt{-b}$ 
whose equation is easily solved:
\begin{eqnarray}
\frac{dh}{dr} \, &=&\, \frac{h^2}{r^2}\sqrt{2}p^3\Rightarrow\nonumber\\
h(r)\, &=&\, H(r)^{-1}\nonumber\\
H(r)\, &=&\,A+\frac{k}{r}\nonumber\\
k\, &=&\,\sqrt{2}p^3
\end{eqnarray}
In what follows we shall put $A=1$.

The equations for the independent axions $a_1,a_2$ may then be rewritten as follows:
\begin{eqnarray}
\frac{da_i}{dr}\, &=&\, \left(\frac{h(r)}{r^2}\right)\sqrt{2}p^3\left(a_i-a_i^{fix}\right)\,\,\,i=1,2
\end{eqnarray}
The solution of the above equations is :
\begin{eqnarray}
\cases{a_i(r)\, =\,\frac{H_i(r)}{H(r)} \cr 
H_i(r)\, =\, \left(a_i^{\infty}+\frac{k_i}{r}\right)\,=\,\frac{k_i}{r} \quad i=1,2 \cr
k_i\, =\,k a_i^{fix}}
\label{ai}
\end{eqnarray} 
As far as the equation for $b$ is concerned, substituting the
expressions of $a_i(r)$ in $F(a_i)$ one obtains an equation for $b(r)$
which is easily solved by introducing a new harmonic function
$H_4(r)$.  Once $b(r)$ is known, from the expression of $h(r)$ it is
straightforward to derive the solution for ${\cal U}(r)$. The result for the
independent evolving fields $a_{1,2}(r),b(r),{\cal U}(r)$ is summarized below
in terms of the four different harmonic functions
$H(r),H_{1,2}(r),H_4(r)$:
\begin{eqnarray}
\cases{a_i(r)\, =\,\frac{H_i(r)}{H(r)} \cr
b(r)\, =\,-\sqrt{\frac{H(r) H_4(r) + m H_1(r) + n H_2(r)}{H(r)^2}} \cr
e^{{\cal U}(r)}\, =\,\left[H(r)^2 (H(r) H_4(r) + m H_1(r) + n H_2(r))\right]^{-1/4} \cr
m = -\frac{q_3}{\alpha p_3}\;\;,\;\; n = \alpha^2 a_2^{fix} \cr
H_4(r)\, =\,1+\frac{k_4}{r}\;\;,\;\; k_4\,=\,k \,b_{fix}^2} 
\label{alltogether}
\end{eqnarray}
It can be shown that the above functional expressions for the fields $a_i(r), b(r)$
and $U(r)$ fulfill the field equations and therefore are a solution of the 
theory. Eq.(\ref{alltogether}), together with eq.s (\ref{charcond})-(\ref{imz0}), represents our 
5 parameter generating solution.
 
Let us consider now the {\it near--horizon} limit of the function $e^{{\cal U}(r)}$:
\begin{eqnarray}
\lim_{r\rightarrow 0}e^{{\cal U}(r)}\,&=&\, r\left(k^3k_4\right)^{-1/4}\nonumber\\
k^3k_4\, &=&\,P_{(4)}(p,q)
\end{eqnarray}   
Substituting this limit in the ans\"atze for the metric (\ref{fishlicense})
we obtain the known result that the near horizon geometry of a regular BPS black hole is described by a 
Bertotti--Robinson metric:
\begin{eqnarray}
ds^2\,&=&\,\frac{r^2}{M_{BR}^2}dt^2-\frac{M_{BR}^2}{r^2}dr^2-M_{BR}^2
\left({\rm sin}^2(\theta)d\phi^2+d \theta^2\right)\nonumber\\
M_{BR}^2\,&=&\,\sqrt{P_{(4)}(p,q)}\nonumber
\end{eqnarray}
The entropy is proportional to the area of the horizon ($Area_H = 4 \pi M^2_{BR}$) according to the 
Bekenstein--Hawking formula:
\begin{equation}
S_{BH}\,=\, \frac{Area_H}{4 G_N}\,\stackrel{\small G_N=1}{=}\,\pi \sqrt{P_{(4)}(p,q)}\,=
\,2 \pi \sqrt{(p^3)^3 q_0 + p^1 q_1 p^3 q_3 - \left(p^1 q_1 + p^3 q_3\right)^2}
\end{equation}
where we have expressed $\alpha$ as the ratio $p^1/p^3$.

As a final remark, we would like to comment on the difference between the form of the generating 
solution in (\ref{alltogether}) and the one described in {\it Statement 3.1} of \cite{bft}, which 
was characterized by two double--fixed axions. The functional expression of the generating solution 
obviously depends on the gauge fixing procedure adopted and, for computational convenience, in the 
present work we have fixed the $[SO(2)]^3$ gauge using a prescription which is different from that 
suggested in \cite{bft}. 

\section{Conclusions}

In this paper we have constructed the generating solution for BPS
saturated static black holes  of $N=8$ supergravity (that is either
M--theory compactified on $T^7$, or equivalently type II  string
theory compactified on $T^6$). This solution preserves 4 supercharges
and can  be seen either as 1/8 preserving in the context of $N=8$
theory or as 1/2 preserving in a type  II Calabi--Yau compactification
(or eventually 1/4 supersymmetry preserving in heterotic on $T^6$ or
type II on  $K^3 \times T^2$). Actually, among compactification of
string theory down to four dimensions, is the only regular  one,
modulo $U$-duality transformation, irrespectively of the original
ten--dimensional theory. In order for it  to act as a generating
solution one has to embed it in the different $U$--duality  groups
($E_{7(7)}$ for toroidal type II compactifications and $SO(6,22)$ for
heterotic ones). 

Our 5 parameter solution should be related via $U$--duality
transformations to that found in \cite{hull}. While both these
solutions carry only NS--NS charges, the proper group  theoretical
embedding of our $STU$ model generating solution in the $N=8$ theory
(\cite{bft})  allows one to obtain, in principle, the macroscopic
description of pure R--R black holes which  can be interpreted
microscopically in terms of D--branes only \cite{bala1}. There has
been an intense  study in giving the precise correspondence between
macroscopic and microscopic black hole configurations  in the last
couple of years. This has been investigated both in the context of
$N=8$ compactifications  (see for example \cite{bala1}-\cite{kleb})
and of $N=2$ compactifications (see for example
\cite{malda}--\cite{dewit}). However, all these solutions were somewhat
particular under one  circumstance or another. What we mean is that a
precise and general recipe to give this  correspondence for {\it any}
macroscopic configuration is still lacking. On the contrary, if we
know how to  transform the generating solution into a generic one, in
particular to those whose microscopic  interpretation is known, then
we can derive the microscopic stringy description of any   geometric
macroscopic solution. And this could shed light even on some still not
understood aspects  of black hole physics as, for example, the very
conceptual basis of the microscopic entropy  counting (for recent work
in this direction see for instance \cite{gibb}). While most of the
group theoretical  machinery necessary in order to make the embedding
has been already constructed in \cite{bft}, it is our aim, in a
forthcoming paper, to give some explicit examples of the use of
$U$--duality transformations in moving from a given  solution to other
ones \cite{bt}. For example, one could find the macroscopic description
corresponding to the five  parameter R--R configuration of \cite{bala}
which, to our knowledge, is the only pure R--R microscopic
configuration  depending on five parameters present in the literature.

\vskip 7pt

{\bf Acknowledgments}

M. B. and M. T. would like to thank  the Theoretical Physics Department of Turin 
University for the kind hospitality during some stages of this work. We also thank 
G. Bonelli, M. S. Gelli and C. Leone for various fruitful discussions. 

\appendix
\section*{Appendix A: the full set of first and second order differential equations}
\label{appendiceB}
\setcounter{equation}{0}
\addtocounter{section}{1}
Setting $z^i=a_i+{\rm i}b_i$ eqs.(\ref{eqs122}) can be rewritten in the form:
{\small
\begin{eqnarray}
\label{mammamia}
\frac{da_1}{dr}\,&=&\,\pm \frac{e^{{\cal U}(r)}}{r^2}\sqrt{- \frac{b_1}{2b_2b_3}}
[-{b_1q_1} + {b_2q_2} + {b_3q_3} +
  \left( -\left( {a_2}\,{a_3}\,{b_1} \right)  +
     {a_1}\,{a_3}\,{b_2} + {a_1}\,{a_2}\,{b_3} +
     {b_1}\,{b_2}\,{b_3} \right) \,{p^0} + \nonumber \\
&&  + \left( -\left( {a_3}\,{b_2} \right)  - {a_2}\,{b_3} \right) \,
   {p^1} + \left( {a_3}\,{b_1} - {a_1}\,{b_3} \right) \,
   {p^2} + \left( {a_2}\,{b_1} - {a_1}\,{b_2} \right) \,
   {p^3}] \nonumber\\
\frac{db_1}{dr}\,&=&\,\pm \frac{e^{{\cal U}(r)}}{r^2}\sqrt{- \frac{b_1}{2b_2b_3}}
[{a_1q_1} + {a_2q_2} + {a_3q_3} +
  \left( {a_1}\,{a_2}\,{a_3} + {a_3}\,{b_1}\,{b_2} +
     {a_2}\,{b_1}\,{b_3} - {a_1}\,{b_2}\,{b_3} \right) \,
   {p^0} + \nonumber \\
&& + \left( -\left( {a_2}\,{a_3} \right)  + {b_2}\,{b_3}
      \right) \,{p^1}
 - \left( {a_1}\,{a_3} + {b_1}\,{b_3} \right)
     \,{p^2} - \left( {a_1}\,{a_2} + {b_1}\,{b_2} \right) \,
   {p^3} + {q_0}] \nonumber\\
\frac{da_2}{dr}\,&=&\, (1,2,3) \rightarrow (2,1,3) \nonumber\\
\frac{db_2}{dr}\,&=&\, (1,2,3) \rightarrow (2,1,3) \nonumber\\
\frac{da_3}{dr}\,&=&\, (1,2,3) \rightarrow (3,2,1) \nonumber\\
\frac{db_3}{dr}\,&=&\, (1,2,3) \rightarrow (3,2,1) \nonumber\\
\frac{d\cal {U}}{dr}\,&=&\, \pm \frac{e^{{\cal U}(r)}}{r^2}\frac{1}{2\sqrt{2}
(- b_1b_2b_3)^{1/2}}[{a_1q_1} + {a_2q_2} + {a_3q_3} +
  \left( {a_1}\,{a_2}\,{a_3} - {a_3}\,{b_1}\,{b_2} -
     {a_2}\,{b_1}\,{b_3} - {a_1}\,{b_2}\,{b_3} \right) \,
   {p^0} + \nonumber \\
&& - \left( {a_2}\,{a_3} - {b_2}\,{b_3} \right) \,
   {p^1} - \left( {a_1}\,{a_3} - {b_1}\,{b_3} \right) \,
   {p^2} - \left( {a_1}\,{a_2} - {b_1}\,{b_2} \right) \,
   {p^3} + {q_0}] \nonumber\\
0\,&=&\,{b_1q_1} + {b_2q_2} + {b_3q_3} +
  \left( {a_2}\,{a_3}\,{b_1} + {a_1}\,{a_3}\,{b_2} +
     {a_1}\,{a_2}\,{b_3} - {b_1}\,{b_2}\,{b_3} \right) \,
   {p^0} - \left( {a_3}\,{b_2} + {a_2}\,{b_3} \right) \,
   {p^1}\nonumber\\
&& - \left( {a_3}\,{b_1} + {a_1}\,{b_3} \right) \,
   {p^2} - \left( {a_2}\,{b_1} + {a_1}\,{b_2} \right) \,
   {p^3}
\end{eqnarray}
}

The explicit form of the equations of motion for the most general case is:
{\small
\begin{eqnarray}
\underline{\mbox{Scalar equations :}}\quad\quad \quad\quad&&\nonumber\\
\left(a_1^{\prime\prime}-2\frac{a_1^{\prime}b_1^{\prime}}{b_1}+2\frac{a_1^{\prime}}{r}\right)\,&=&\,
\frac{-2\,{b_1}\,{e^{2\,U}}}{{r^4}}\,
     [ {a_1}\,{b_2}\,{b_3}\,( {{{p^0}}^2} - {{{\ell (r)_0}}^2} ) +
       {b_2}\,( -( {b_3}\,{p^0}\,{p^1} )
              + {b_3}\,{\ell (r)_0}\,{\ell (r)_1} )  + \nonumber \\
&&      + {b_1}\,( -2\,{a_2}\,{a_3}\,{p^0}\,
           {\ell (r)_0} + {a_3}\,{p^2}\,{\ell (r)_0} +
           {a_2}\,{p^3}\,{\ell (r)_0} +
          {a_3}\,{p^0}\,{\ell (r)_2} + \nonumber \\
&&        -  {p^3}\,{\ell (r)_2} +
          {a_2}\,{p^0}\,{\ell (r)_3} - {p^2}\,{\ell (r)_3}
           )  ]  \nonumber\\
\left(b_1^{\prime\prime}+2\frac{b_1^{\prime }}{r}+
\frac{(a_1^{\prime 2}-b_1^{\prime2})}{b_1}\right)\,&=&\,
 -\frac{{e^{2\,U}}}{{b_2}\,{b_3}\,{r^4}}\,[ -( {{{a_1}}^2}\,{{{b_2}}^2}\,
           {{{b_3}}^2}\,{{{p^0}}^2} )  +
        {{{b_1}}^2}\,{{{b_2}}^2}\,{{{b_3}}^2}\,
         {{{p^0}}^2} + 2\,{a_1}\,{{{b_2}}^2}\,
         {{{b_3}}^2}\,{p^0}\,{p^1} + \nonumber \\
&&       - {{{b_2}}^2}\,{{{b_3}}^2}\,{{{p^1}}^2} +
        {{{b_1}}^2}\,{{{b_3}}^2}\,{{{p^2}}^2} +
        {{{b_1}}^2}\,{{{b_2}}^2}\,{{{p^3}}^2} +
        {{{a_1}}^2}\,{{{b_2}}^2}\,{{{b_3}}^2}\,
         {{{\ell (r)_0}}^2} + \nonumber \\
&&       - {{{b_1}}^2}\,{{{b_2}}^2}\,
         {{{b_3}}^2}\,{{{\ell (r)_0}}^2} +
        {{{a_3}}^2}\,{{{b_1}}^2}\,{{{b_2}}^2}\,
         ( {{{p^0}}^2} - {{{\ell (r)_0}}^2} )  +
        {{{a_2}}^2}\,{{{b_1}}^2}\,{{{b_3}}^2}\nonumber \\
&&         ( {{{p^0}}^2} - {{{\ell (r)_0}}^2} )
        - 2\,{a_1}\,{{{b_2}}^2}\,{{{b_3}}^2}\,{\ell (r)_0}\,
         {\ell (r)_1} +
          {{{b_2}}^2}\,{{{b_3}}^2}\,
         {{{\ell (r)_1}}^2} + \nonumber \\
&&       - {{{b_1}}^2}\,{{{b_3}}^2}\,
         {{{\ell (r)_2}}^2} +
          2\,{a_2}\,{{{b_1}}^2}\,
         {{{b_3}}^2}\,( -( {p^0}\,{p^2} )  +
           {\ell (r)_0}\,{\ell (r)_2} ) + \nonumber \\
&&       - {{{b_1}}^2}\,{{{b_2}}^2}\,{{{\ell (r)_3}}^2}
       + 2\,{a_3}\,{{{b_1}}^2}\,{{{b_2}}^2}\,
         ( -( {p^0}\,{p^3} )  +
           {\ell (r)_0}\,{\ell (r)_3} )  ] \,
      \nonumber\\
\left(a_2^{\prime\prime}-2\frac{a_2^{\prime}b_2^{\prime}}{b_2}+2\frac{a_2^{\prime}}{r}\right)\,&=&\,
 (1,2,3) \rightarrow (2,1,3) \nonumber\\
\left(b_2^{\prime\prime}+2\frac{b_2^{\prime }}{r}+
\frac{(a_2^{\prime
2}-b_2^{\prime2})}{b_2}\right)\,&=&\,
 (1,2,3) \rightarrow (2,1,3)  \nonumber\\
\left(a_3^{\prime\prime}-2\frac{a_3^{\prime}b_3^{\prime}}{b_3}+2\frac{a_3^{\prime}}{r}\right)\,&=&\,
 (1,2,3) \rightarrow (3,2,1)    \nonumber\\
\left(b_3^{\prime\prime}+2\frac{b_3^{\prime }}{r}+
\frac{(a_3^{\prime
2}-b_3^{\prime2})}{b_3}\right)\,&=&\,
 (1,2,3) \rightarrow (3,2,1) \nonumber \\
\underline{\mbox{Einstein equations :}}\quad\quad\quad\quad &&\nonumber\\
{\cal U}^{\prime\prime}+\frac{2}{r}{\cal U}^\prime\,&=&\,-2e^{-2{\cal U}}
S_{00} \nonumber\\
({\cal U}^\prime)^2+\sum_i\frac{1}{4b_i^2}\left((b_i^\prime)^2+(a_i^\prime)^2\right)\,&=&\,
-2e^{-2{\cal U}}S_{00}
\label{porcodue}
\end{eqnarray}
}
where the quantity $S_{00}$ on the right hand side of the Einstein eqs.
has the following form:
{\small
\begin{eqnarray}
S_{00}\, &=&\,\frac{{e^{4\,U}}}{4\,{b_1}\,{b_2}\,{b_3}\,{r^4}}\,
( {{{a_1}}^2}\,{{{b_2}}^2}\,{{{b_3}}^2}\,
        {{{p^0}}^2} + {{{b_1}}^2}\,{{{b_2}}^2}\,
        {{{b_3}}^2}\,{{{p^0}}^2} -
       2\,{a_1}\,{{{b_2}}^2}\,{{{b_3}}^2}\,{p^0}\,
        {p^1} + {{{b_2}}^2}\,{{{b_3}}^2}\,{{{p^1}}^2} +
       {{{b_1}}^2}\,{{{b_3}}^2}\,{{{p^2}}^2} + \nonumber \\
&&      + {{{b_1}}^2}\,{{{b_2}}^2}\,{{{p^3}}^2} +
       {{{a_1}}^2}\,{{{b_2}}^2}\,{{{b_3}}^2}\,
        {{{\ell (r)_0}}^2} + {{{b_1}}^2}\,{{{b_2}}^2}\,
        {{{b_3}}^2}\,{{{\ell (r)_0}}^2} +
       {{{a_3}}^2}\,{{{b_1}}^2}\,{{{b_2}}^2}\,
        ( {{{p^0}}^2} + {{{\ell (r)_0}}^2} ) + \nonumber \\
&&     + {{{a_2}}^2}\,{{{b_1}}^2}\,{{{b_3}}^2}\,
        ( {{{p^0}}^2} + {{{\ell (r)_0}}^2} )  -
       2\,{a_1}\,{{{b_2}}^2}\,{{{b_3}}^2}\,{\ell (r)_0}\,
        {\ell (r)_1} + {{{b_2}}^2}\,{{{b_3}}^2}\,
        {{{\ell (r)_1}}^2} + {{{b_1}}^2}\,{{{b_3}}^2}\,
        {{{\ell (r)_2}}^2} + \nonumber \\
&&       - 2\,{a_2}\,{{{b_1}}^2}\,
        {{{b_3}}^2}\,( {p^0}\,{p^2} +
          {\ell (r)_0}\,{\ell (r)_2} )  +
       {{{b_1}}^2}\,{{{b_2}}^2}\,{{{\ell (r)_3}}^2} -
       2\,{a_3}\,{{{b_1}}^2}\,{{{b_2}}^2}\,
        ( {p^0}\,{p^3} + {\ell (r)_0}\,{\ell (r)_3} ))
\end{eqnarray}
}
The explicit expression of the $\ell_{\Lambda}(r)$ charges in terms of the quantized ones is:
\begin{equation}
\hskip -3pt \ell_{\Lambda} (r) \,=\,\left(\matrix{{\frac{{q_0} + {a_1}\,
      \left( {a_2}\,{a_3}\,{p^0} -
        {a_3}\,{p^2} - {a_2}\,{p^3} + {q_1}
         \right)  + {a_2}\,\left( -\left( {a_3}\,{p^1} \right)
            + {q_2} \right)  + {a_3}\,{q_3}}{{b_1}\,
     {b_2}\,{b_3}}}\cr
{\frac{{{{a_1}}^2}\,\left( {a_2}\,{a_3}\,{p^0} -
        {a_3}\,{p^2} - {a_2}\,{p^3} + {q_1}
         \right)  + {{{b_1}}^2}\,
      \left( {a_2}\,{a_3}\,{p^0} -
        {a_3}\,{p^2} - {a_2}\,{p^3} + {q_1}
         \right)  + {a_1}\,\left( {q_0} +
        {a_2}\,\left( -\left( {a_3}\,{p^1} \right)  +
           {q_2} \right)  + {a_3}\,{q_3} \right) }{
     {b_1}\,{b_2}\,{b_3}}}\cr
{\frac{{a_1}\,\left( {{{a_2}}^2}\,
         \left( {a_3}\,{p^0} - {p^3} \right)  +
        {{{b_2}}^2}\,\left( {a_3}\,{p^0} - {p^3}
           \right)  + {a_2}\,
         \left( -\left( {a_3}\,{p^2} \right)  + {q_1} \right)
         \right)  + {{{a_2}}^2}\,
      \left( -\left( {a_3}\,{p^1} \right)  + {q_2} \right)  +
     {{{b_2}}^2}\,\left( -\left( {a_3}\,{p^1} \right)  +
        {q_2} \right)  + {a_2}\,
      \left( {q_0} + {a_3}\,{q_3} \right) }{{b_1}\,
     {b_2}\,{b_3}}}\cr
 {\frac{{a_3}\,{q_0} +
     {a_1}\,\left( -\left( {{{a_3}}^2}\,{p^2} \right)  -
        {{{b_3}}^2}\,{p^2} +
        {a_2}\,\left( {{{a_3}}^2}\,{p^0} +
           {{{b_3}}^2}\,{p^0} - {a_3}\,{p^3} \right)  +
         {a_3}\,{q_1} \right)  -
     {a_2}\,\left( {{{a_3}}^2}\,{p^1} +
        {{{b_3}}^2}\,{p^1} - {a_3}\,{q_2} \right)  +
     {{{a_3}}^2}\,{q_3} + {{{b_3}}^2}\,{q_3}}{
     {b_1}\,{b_2}\,{b_3}}}} \right)
\end{equation}

\end{document}